\documentclass[aps, preprint]{revtex4} % add twocolumn option
\usepackage[utf8]{inputenc}
\usepackage{graphicx}
\usepackage{amssymb}
\usepackage{epstopdf}
\usepackage{cancel}
\usepackage{physics}
\usepackage{hyperref}
\usepackage{xcolor}

\newcommand{\Tobs}{T_{\rm obs}}

\newcommand{\refeq}[1]{Eq.~\eqref{eq:#1}}

\makeatletter
\g@addto@macro\bfseries{\boldmath}
\makeatother

\begin{document}
\preprint{} % MPP preprint code
\title{Is the bump significant? An axion-search example}
\author{Frederik Beaujean}
\affiliation{C2PAP, Excellence Cluster Universe, Ludwig-Maximilian University of Munich}
\author{Allen Caldwell}
\author{Olaf Reimann}
\affiliation{Max Planck Institute for Physics, Munich}
\date{\today}                                           % Activate to display a given date or no date

\begin{abstract}
  Many experiments in physics involve searching for a localized excess over background expectations in an observed spectrum. If the background is known and there is Gaussian noise, the amount of excess of successive observations can be quantified by the runs statistic taking care of the look-elsewhere effect. The distribution of the runs statistic under the background model is known analytically but the computation becomes too expensive for more than about a hundred observations. This work demonstrates a principled high-precision extrapolation from a few dozen up to millions of data points. It is most precise in the interesting regime when an excess is present. The method is verified for benchmark cases and successfully applied to real data from an axion search. The code that implements our method is available at https://github.com/fredRos/runs.
\end{abstract}

\maketitle

\section{Introduction}\label{sec:intro}

We revisit the problem of searching for a bump at an unknown location in a
spectrum. Specifically we assume there are $L$ observations $\qty{y_i}$ and the
index $i$ provides an ordering for the data, for example time, mass,
energy\dots. In our background model that has no bump, the observations
independently follow a Gaussian or Normal distribution
\begin{align}
  \label{eq:normaldata}
  y_i \sim \mathcal{N}\qty(\mu_i, \sigma_i)
\end{align}
where the expectation $\mu$ and the standard deviation $\sigma$ are known for
every $i$. In our previous work~\cite{beaujean2011test}, we introduced the runs test
statistic $T$ to check the consistency of the background model with the
observations. If a discrepancy is found, more specific analyses can be carried
out to decide if a signal is present and to determine the parameters of the signal.

The main motivation behind the runs statistic is that it automatically takes
care of the look-elsewhere effect (also called the trials factor) that arises in
some other methods that look for a narrow peaks, for example in the search for
the Higgs boson at the LHC~\cite{Aad:2012tfa,Chatrchyan:2012xdj}. There, the
profile-likelihood ratio statistic was employed~\cite{Cowan:2010js} which
requires fully specifying both the background and the signal model including
dependence on unknown parameters to be estimated from the data. For reliable
estimates of the look-elsewhere effect, asymptotic normality and principled
extrapolation from small to large significance had to be
used~\cite{Vitells:2011bba,Gross:2010qma}.

In comparison, the runs statistic does not require a signal model and does not
rely on asymptotic normality of the likelihood but assumes the background is
known exactly. In \cite{beaujean2011test}, it was demonstrated that in this
setting the runs statistic leads to a more powerful test than the classic
$\chi^2$ test in this peak-fitting problem.

Recounting the definition of the runs statistic, consider the sequence of $L$
observations as consisting of success and failure runs, where the observation $i$ is
a success if it is above the background expectation, $y_i \geq \mu_i$. The runs statistic
$T$ is defined as the largest value of $\chi^2$ for any success run
\begin{equation}
T \equiv \max_R \sum_{i\in R} \left(\frac{y_i -\mu_i}{\sigma_i}\right)^2,
\end{equation}
where $R$ represents the set of indices in an individual success run. Using the cumulative $F(T | L)$, the $p$ value is the tail-area probability to find $T$ larger than the observed value $\Tobs$,
\begin{align}
  \label{eq:pvalue}
  p \equiv 1-F(\Tobs | L) \,.
\end{align}

\begin{figure}[htbp]
   \centering
  \includegraphics[width=4in]{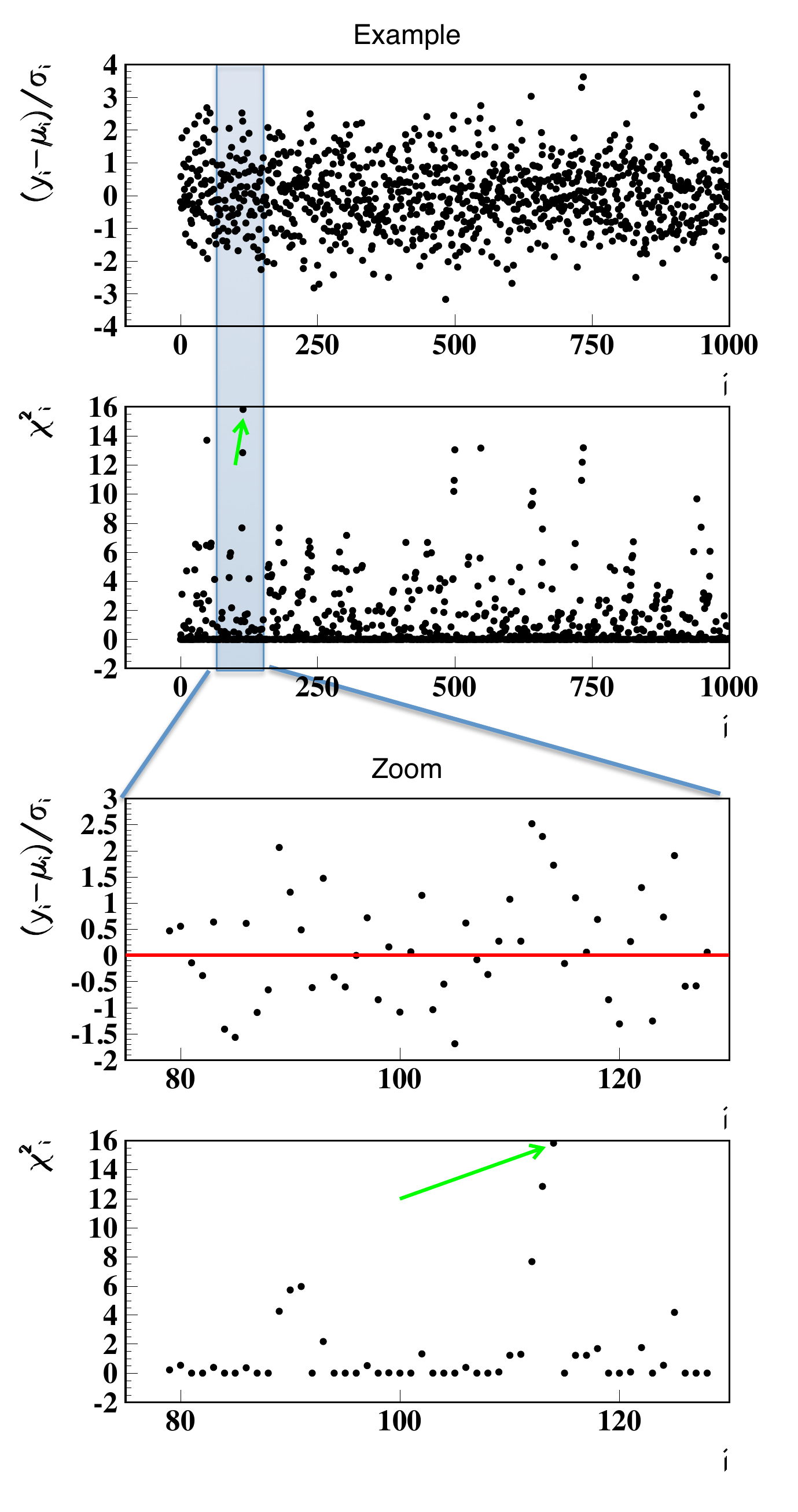}
  \caption{A sequence of $1000$ Gauss distributed random numbers (shifted and
    scaled) is shown in the top plot, while the value $\chi^2_i$ of the current
    success run is shown in the second plot. A zoom in the region around the
    largest $\chi^2_i$ is shown below.
}
 \label{fig:onerun}
\end{figure}

As an illustration, a sequence of $L=1000$ independent standard Gauss distributed random
numbers (shifted and scaled) is shown in Fig.~\ref{fig:onerun}, while
the running value of $\chi^2$, $\chi^2_i$, is shown in the second
plot. Note that $\chi^2_i$ is $0$ initially and as soon as a failure
is encountered; i.e., $y_i<\mu_i$ . Otherwise, it is incremented by
$(y_i-\mu_i)^2/\sigma_i^2$ for every success. In the case shown, the
largest observed $\chi^2$ of any run leads to $T_{\rm
  obs}=15.8$. Using our results presented below, this $\Tobs$ is
equivalent to a $p$ value of 0.36 which suggests good agreement with
the background-only hypothesis.

The exact probability distribution for the runs statistic $T$ has been derived in
~\cite{beaujean2011test}, and code is available on
github~\cite{frederik_beaujean_2017_845743} to calculate the cumulative of the
test statistic in \texttt{mathematica} and \texttt{C++}. The calculation time
grows rapidly with $L$ (roughly as $\exp(\sqrt{L})/L$), and for $L \approx 200$
becomes too long for practical use on todays CPUs, even with multiple cores. We
derive here a formula that allows us to use the results for moderate $L \approx
100$ to extrapolate to very large $L \gtrapprox 10^6$ with high accuracy in the
region of interest where $\Tobs$ is large such that the $p$ value is very small.
We have implemented the extrapolation formula in our
code~\cite{frederik_beaujean_2017_845743}. This allows the use of the runs
statistic in very long sequences of measurements with the correct statistical
distribution for the test statistic without relying on expensive and somewhat
inaccurate Monte Carlo simulations.

An application of our run statistic is described in the last section of this paper.  The setting is an axion search experiment~\cite{TheMADMAXWorkingGroup:2016hpc}, where we eventually expect to have of order $5 \cdot 10^7$ power measurements integrated over $\sim 2$~kHz intervals covering a frequency range of approximately $100$~GHz.  The data will be acquired in approximately $50$~MHz data sets.  The axion signal is expected to be very narrow, possible one to several $2$~kHz bins wide, but with unknown shape, and we wish to make the minimum number of assumptions in a first pass through the data.  We intend to use the run statistic to identify candidate signals in this spectrum.  Once a candidate signal is identified, the experimental setup can be modified to increase the signal-to-noise ratio considerably.  However, changing the setup and acquiring more data is time consuming, and can only be performed relatively rarely.  It is therefore important to understand the statistical significance of a putative signal.

\section{Cumulative of the test statistic for large $L$}

We assume there is a sequence of $L$ observations. Consider a partition
$L=N_l+N_r$ into two segments, where $N_l$ and $N_r$ denote the left-hand and
right-hand part. Suppose that in each of the segments considered separately, the
test statistics $T_l$ and $T_r$ are both less than $\Tobs$. Then there are
exactly two ways this can occur. Either $T < \Tobs$ for the entire sequence, or
there is a run that crosses the boundary (condition $\mathcal{C}$) and its
$\chi^2$ in each segment is less than $\Tobs$ but the combined $\chi^2$ is the
largest of any run (condition $\mathcal{M}$) and exceeds $\Tobs$:
\begin{align}
  \label{eq:mutual-events}
  P(T_l<\Tobs, T_r<\Tobs | N_l, N_r, L) = &P(T<\Tobs | N_l, N_r, L) + \nonumber\\
                                          &P(T_l<T_{\rm obs}, T_r<T_{\rm obs}, T \geq T_{\rm obs}, \mathcal{C,M} | N_l, N_r, L)
\end{align}
We denote by $F(\Tobs|L) \equiv P(T<\Tobs | \cancel{N_l}, \cancel{N_r}, L)$ the value of the cumulative
probability for the test statistic $T$ for a total of $L$ observations.
Since the events are assumed independent, we can factorize
\begin{align}
  P(T_l<\Tobs, T_r<\Tobs | N_l, N_r, L) &= P(T_l<\Tobs | N_l, \cancel{N_r}, \cancel{L}) P(T_r<\Tobs | \cancel{N_l}, N_r, \cancel{L}) \\
  &=F(\Tobs | N_l)F(\Tobs | N_r)
\end{align}
into the left and right parts and rearrange to find
% Consider a partition $L=N_l+N_r$ into two segments, where $N_l$ and $N_r$ denote the left-hand and right-hand part.  The probability that the test statistic is less than an observed value $T_{\rm obs}$ is the probability that it is below $T_{\rm obs}$ in both the $N_l$ segment and the $N_r$ segment, corrected for a contribution where runs across the boundary between the two segments, and where this run satisfies the condition on each side but has $T>T_{\rm obs}$ and this is the largest run.
% I.e.,
\begin{equation}
  \label{eq:decomp}
F(T_{\rm obs}|L) = F(T_{\rm obs}|N_l)F(T_{\rm obs}|N_r) - P(T_l<T_{\rm obs}, T_r<T_{\rm obs}, T \geq T_{\rm obs}, \mathcal{C, M} | N_l, N_r, L) \; .
\end{equation}
For both $N_l$ and $N_r$ large, there typically are many runs so it is unlikely
that a boundary-spanning run has the largest $\chi^2$. Then
\refeq{decomp} shows that the cumulative of the whole sequence is essentially the
product of cumulatives for each segment minus a small correction.
\begin{figure}[htbp]
   \centering
\includegraphics[width=\textwidth]{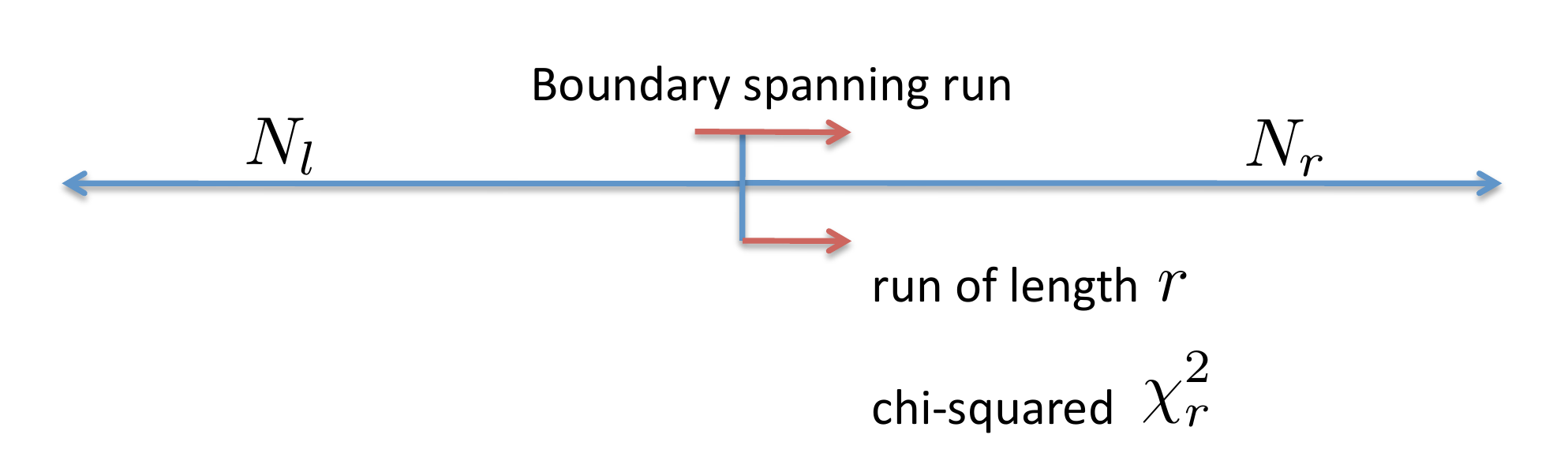}
  \caption{Visual description of reducing a long run into two shorter runs.
}
 \label{fig:split}
\end{figure}

\subsection{$\chi^2$ distribution of runs starting at the boundary}

Since the boundary is fixed and we require for the correction term that the run cross the boundary, it must necessarily have the first result on each side above the expectation. Consider the run segment on the right of the boundary: we can use the law of total probability to calculate the  probability density as the sum of probability densities for runs of different length times the $\chi^2$ probability for that number of degrees of freedom:
$$h(\chi^2|N_r) = \sum_{r=1}^{N_r} P(\chi^2|r) P(r|N_r)$$
where $r$ is the length of the success run and we require at least one success.
We have $P(r|N_r)=(1/2)^{r+1}$ for $r<N_r$ and $P(r=N_r|N_r)=(1/2)^{N_r}$.  The reason for the $+1$ for $r<N_r$ is the requirement that the result is below expectation for the $r+1^{\rm st}$ sample, which also has probability $1/2$. $P(\chi^2|r)$ is the usual chi-squared probability density for $r$ degrees of freedom so that we find
\begin{align}
\label{eq:PT2}
h(\chi^2|N_r) = \left(\sum_{r=1}^{N_r-1} (1/2)^{r+1} \frac{(\chi^2)^{r/2-1} e^{-\chi^2/2}}{2^{r/2}\Gamma(r/2)} \right) + (1/2)^{N_r} \frac{(\chi^2)^{N_r/2-1} e^{-\chi^2/2}}{2^{N_r/2}\Gamma(N_r/2)} \,.
\end{align}
For a $\chi^2$ from a particular run to be our test statistic $T$, the run must
span the boundary (condition $\mathcal{C}$) and its $\chi^2$ must be the maximum
value for any run in the $L$ range (condition $\mathcal{M}$), so $\chi^2_l + \chi^2_r=T$.  We calculate the probability that a contiguous run spanning the boundary satisfies the conditions specified as
\begin{align}
\label{eq:PT}
P(\chi^2_{\rm l}<T_{\rm obs}&, \chi^2_{\rm r}<T_{\rm obs}, \chi^2_{\rm l}+\chi^2_{\rm r} \geq T_{\rm obs}, \mathcal{C,M} | N_l, N_r, L) \\
 &=\int_{0}^{T_{\rm obs}}  \dd{\chi^2_{\rm l}}h(\chi^2_{\rm l}|N_l) \int_{T_{\rm obs}-\chi^2_{\rm l}}^{T_{\rm obs}} \dd{\chi^2_{\rm r}} F(\chi^2_l + \chi^2_r|L)h(\chi^2_{\rm r}|N_r) \,. \nonumber
\end{align}
We implement condition $\mathcal{M}$ by weighting each possible contribution
$\chi^2_l+\chi^2_r$ with the probability that this is the largest in the full
range $L$, $F(\chi^2_l + \chi^2_r|L)$. This is the quantity we seek to compute so we
cannot evaluate this expression directly. But
\begin{align}
\label{eq:approx}
F(\Tobs|L) \leq F(\chi^2_l+\chi^2_r|L) \leq   F(2\,T_{\rm obs}|L) \leq 1
\end{align}
so as $F(T_{\rm obs}|L) \rightarrow 1$ we can write
\begin{align}
  P(\chi^2_{\rm l}<T_{\rm obs}&, \chi^2_{\rm r}<T_{\rm obs}, \chi^2_{\rm l}+\chi^2_{\rm r} \geq T_{\rm obs}, \mathcal{C,M} | N_l, N_r, L)\\
  &\equiv x(\Tobs, L) \Delta(T_{\rm obs} | N_l, N_r)                                                                                                   \gtrapprox F(T_{\rm obs}|L)  \Delta(T_{\rm obs} | N_l, N_r)
\end{align}
where we define
\begin{align}
  \label{eq:DeltaN}
  \Delta(\Tobs | N_l, N_r) &\equiv \int_{0}^{T_{\rm obs}}  \dd{\chi^2_{\rm l}}h(\chi^2_{\rm l}|N_l) \eval{H(\chi^2_{\rm r}| N_r)}_{\Tobs - \chi^2_{\rm l}}^{\Tobs}
\end{align}
and $H(\chi^2_{\rm r} | N_r)$ is the cumulative of $h$. $H$ can be expressed in terms of the cumulative of $P(\chi^2 | r)$, and requires no numerical integral.

\noindent
With the approximation $x(\Tobs, L) = F(\Tobs|L)$, \refeq{decomp} now simplifies to
\begin{equation}
    \label{eq:Fdecomp-sum}
F(T_{\rm obs}|L) = F(T_{\rm obs}|N_l)F(T_{\rm obs}|N_r)- F(T_{\rm obs}|L) \Delta(T_{\rm obs}|N_l, N_r)
\end{equation}
or
\begin{equation}
  \label{eq:Fdecomp-frac}
F(T_{\rm obs}|L) = \frac{F(T_{\rm obs}|N_l)F(T_{\rm obs}|N_r)}{1+\Delta(T_{\rm obs}|N_l, N_r)} \; .
\end{equation}

We expect this expression to become exact as $F(T_{\rm obs}|L) \rightarrow 1$,
and to show some discrepancies at smaller values of $F(T_{\rm obs}|L)$ where the
approximation employed in \refeq{approx} is not valid. Since we underestimate
the correction term, we overestimate $F(\Tobs|L)$. The error is larger at values
of $T_{\rm obs}$ where $F$ is finite but not close to $1$. We evaluate this
effect for some examples below. For the more interesting region where $F
\rightarrow 1$, we expect our approximation to be excellent.

The correction $\Delta(T_{\rm obs}|N_l, N_r)$ is nearly independent of $N_l,
N_r$ in our region of interest ($N_l$ and $N_r$ large) due to the Bernoulli
suppression of long runs in $h$ which is $\propto 2^{N_{l,r}/2}$ ; cf.
\refeq{PT2}. We provide numerical results supporting this claim in
Sec.~\ref{sec:numerical}.

Let us consider the special case $N_l=N_r=N$ and assume that indeed
$\Delta(\Tobs | N, N)$ is independent of $N$, so
\begin{align}
  \label{eq:res2N}
  F(T_{\rm obs}|2N) = \frac{F(T_{\rm obs}|N)^2}{1+\Delta(T_{\rm obs})} \; .
\end{align}
Taking $n=2$ as the base case, we can generalize to arbitrary $n \geq 2$ by
induction to arrive at our main result
\begin{equation}
\label{eq:FTnN}
F(T_{\rm obs}|nN) = \frac{F(T_{\rm obs}|N)^n}{(1+\Delta(\Tobs))^{n-1}} \; , n \geq 2.
\end{equation}
To highlight the hidden assumptions in this procedure, we
write down the induction step from $n \to n+1$ in detail using $N_l=nN, N_r=N$ and suppressing the $\Tobs$ dependencies
\begin{align}
  \label{eq:ntonplus1}
  F(\cdot | (n+1)N) &= F(\cdot | nN) F(\cdot | N) - x(\cdot, (n+1)N) \Delta\\
  &= F(\cdot | nN) F(\cdot | N) - F(\cdot | (n+1)N) \Delta\\
  &= \frac{F(\cdot|N)^{n+1}}{(1+\Delta)^{n-1}} - F(\cdot | (n+1)N) \Delta\\
  &= \frac{F(\cdot|N)^{n+1}}{(1+\Delta)^{n}} \, .
\end{align}
This implies that the approximation $x(\Tobs, nN) = F(\Tobs | nN)$ is
consistently employed $n-1$ times and that we neglect contributions from runs
longer than $2N$ which is acceptable for the same reasons that $\Delta$ can be
considered independent of $N$. For concreteness, in our tests we set $\Delta \equiv
\Delta(\Tobs | N) \equiv \Delta(\Tobs | N,N)$.

With this scaling equation, we can use exact results for moderate values of $N$ to find the $p$ value for our $T_{\rm obs}$ for very large $n N$.

\subsection{Trials factor}\label{sec:trials}

In many applications, the look-elsewhere effect just amounts to multiplying the $p$
value by the number of trials, or trials factor. For example in a finely binned
histogram with $n$ bins the $p$ value for the entire histogram is just $n$ times
the largest $p$ value of any single bin \cite{Ranucci:2011chu}. Following the
histogram analogy, we consider a batch of $N$ successive data points as one bin.
Neglecting the denominator in \refeq{FTnN}, the overall $p$ value
\begin{align}
  \label{eq:trials-overall}
   p(\Tobs | nN) \approx 1-  \qty(1-p(\Tobs | N))^n \approx n\, p(\Tobs | N)
\end{align}
if both $p(\Tobs | N)$ and $\Delta$ are small. We identify the trials factor as
the number of batches $n$ and remark that the proportionality is only
approximate in our application as it mildly depends on $\Tobs$ and possibly $N$.

\section{Numerical tests}
\label{sec:numerical}

\begin{figure}[t] %  figure placement: here, top, bottom, or page
   \centering
  \includegraphics[width=5in]{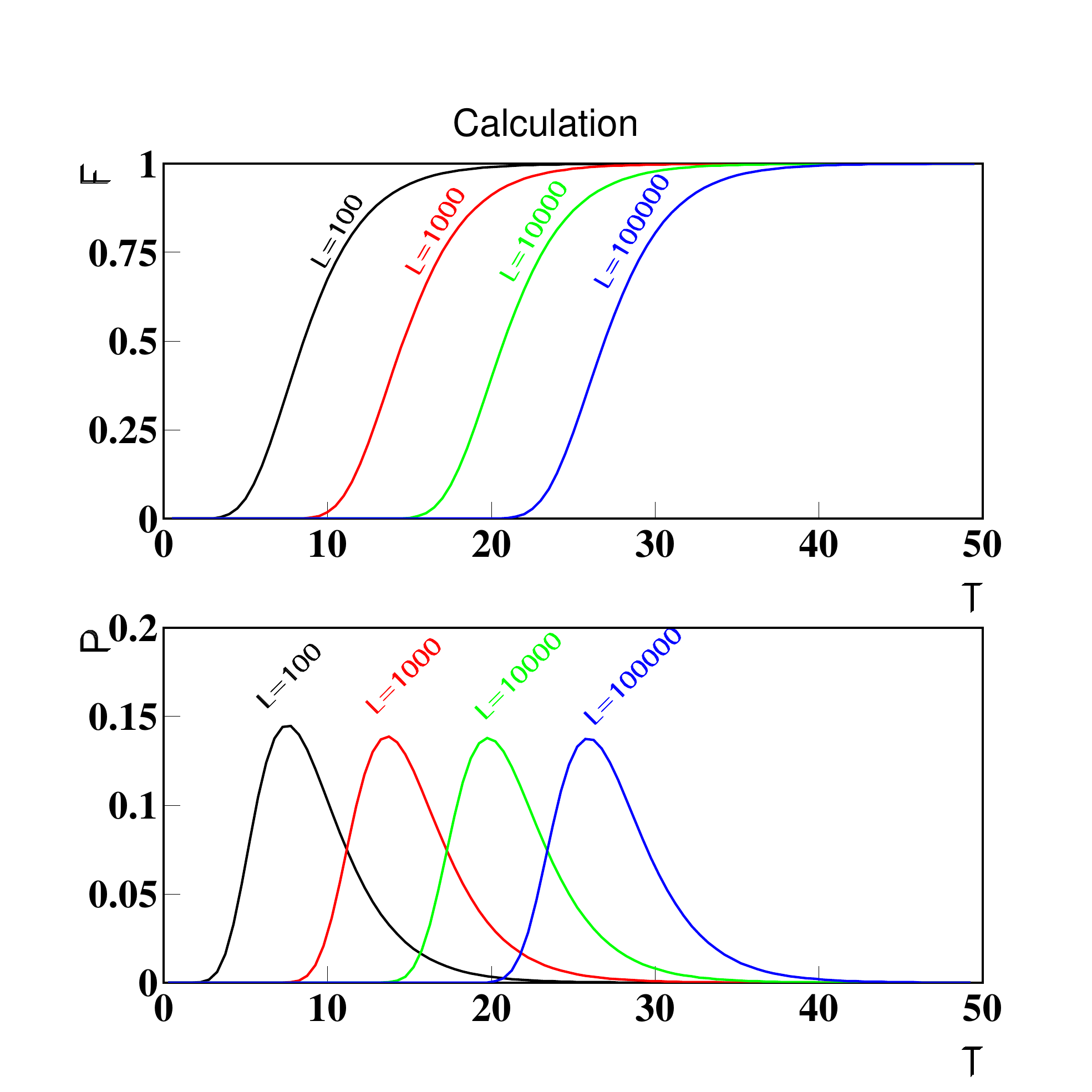}
  \caption{Top: Cumulative probability distributions, $F(T|L)$, for $L=10^2,10^3,10^4,10^5$.  Bottom: Probability density, $P(T|L)$, for the same $L$.}
   \label{fig:Nscaling}
\end{figure}

We first display the behavior of $F(T|L)$ and $P(T|L)$ for different $L$ in Fig.~\ref{fig:Nscaling}.  For $L=100$, the exact calculation is used, while for larger $L$ the scaling formula Eq.~\ref{eq:FTnN} is used. As is seen in the figure, the shape of the probability distribution does not change very much, but only shifts to larger values of $T$ with a speed approximately proportional to $\ln(L)$.

\subsection{$\Delta$ Dependence on $N$}

\begin{figure}[htbp] %  figure placement: here, top, bottom, or page
   \centering
  \includegraphics[width=4in]{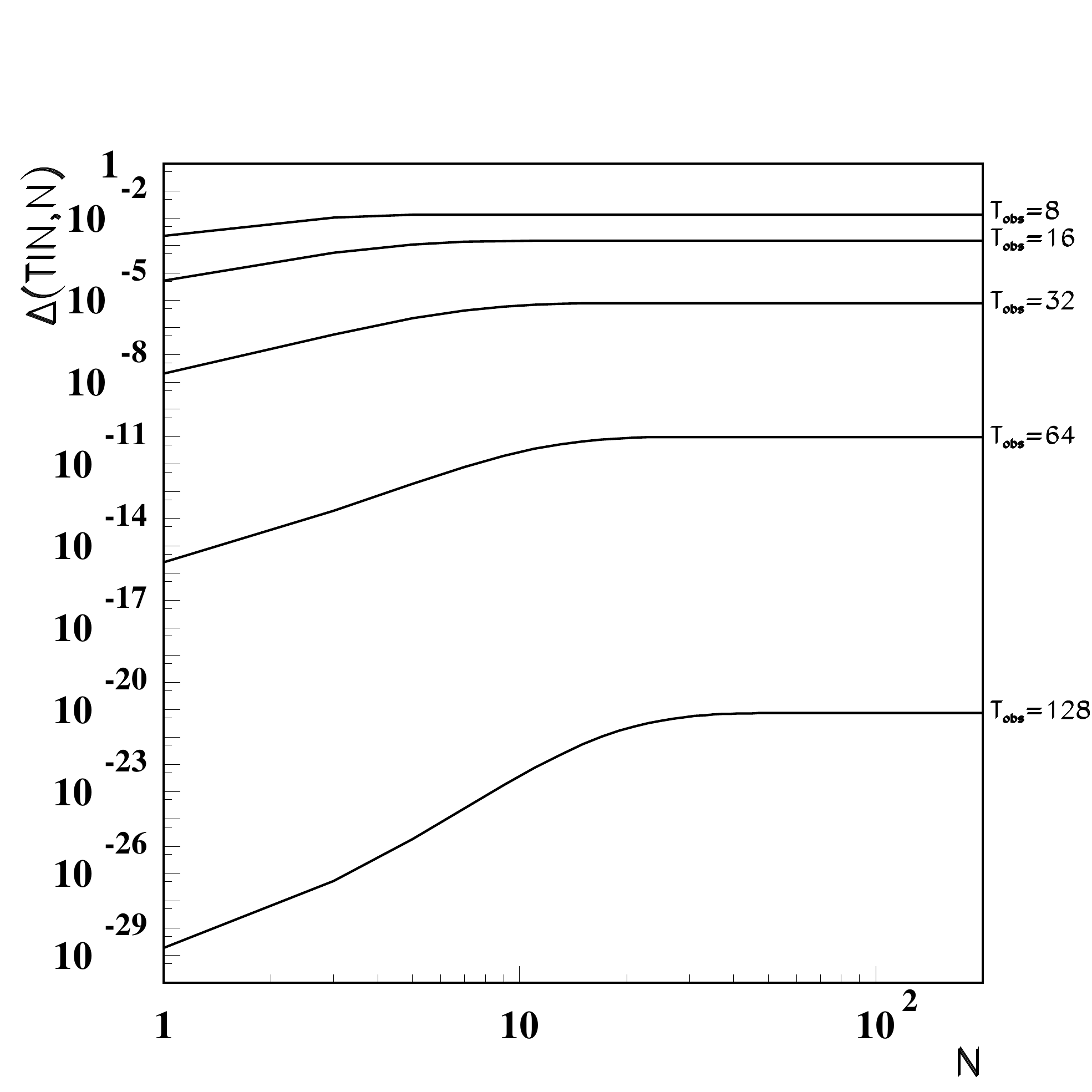}
  \caption{$\Delta(T|N,N)$ as a function of $N$ for several values of $T$.}
 \label{fig:Delta}
\end{figure}

We now show that the quantity $\Delta(T|N)$ is indeed independent of $N$ at
large enough $N$. Figure~\ref{fig:Delta} compares $\Delta(T|N)$ for several
values of $T$. When $T$ is large, $\Delta(T|N)$ is tiny and so is the $p$ value.
While $\Delta(T|N)$ indeed shows variations with $N$ at small $N$, the
dependence on $N$ is negligible at larger $N$. For example for $T=512$,
$\Delta(T|N)$ is well into the saturated region at $N=100$ and
$\Delta(512|100,100) = 10^{-84}$. Since we propose to use our approximation for
$N\gtrsim 100$, the error introduced by ignoring the $N$ dependence of $\Delta$ is
completely negligible.

\subsection{Accuracy of the approximation}

\begin{figure}[thbp] %  figure placement: here, top, bottom, or page
   \centering
  \includegraphics[width=2.5in]{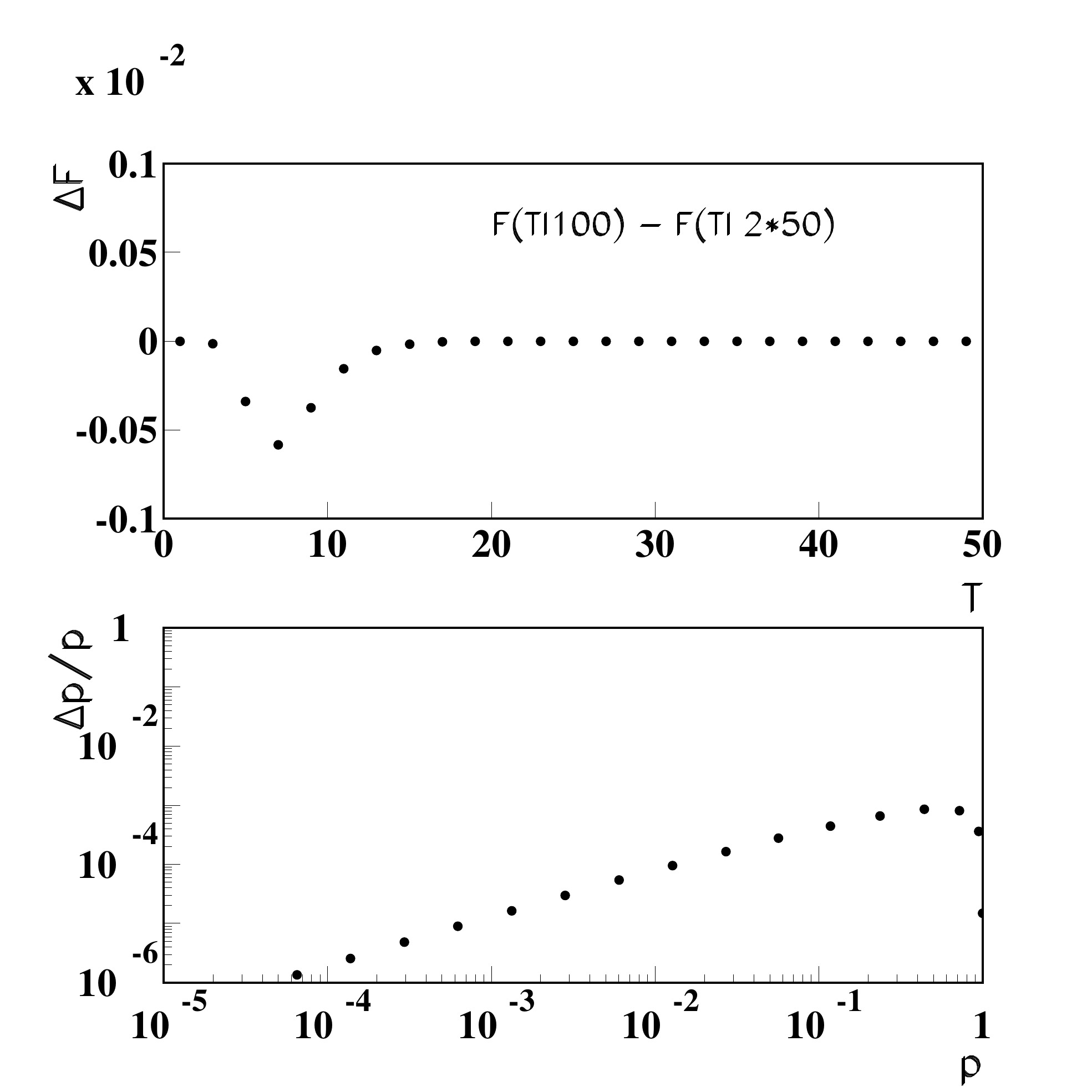}
  \includegraphics[width=2.5in]{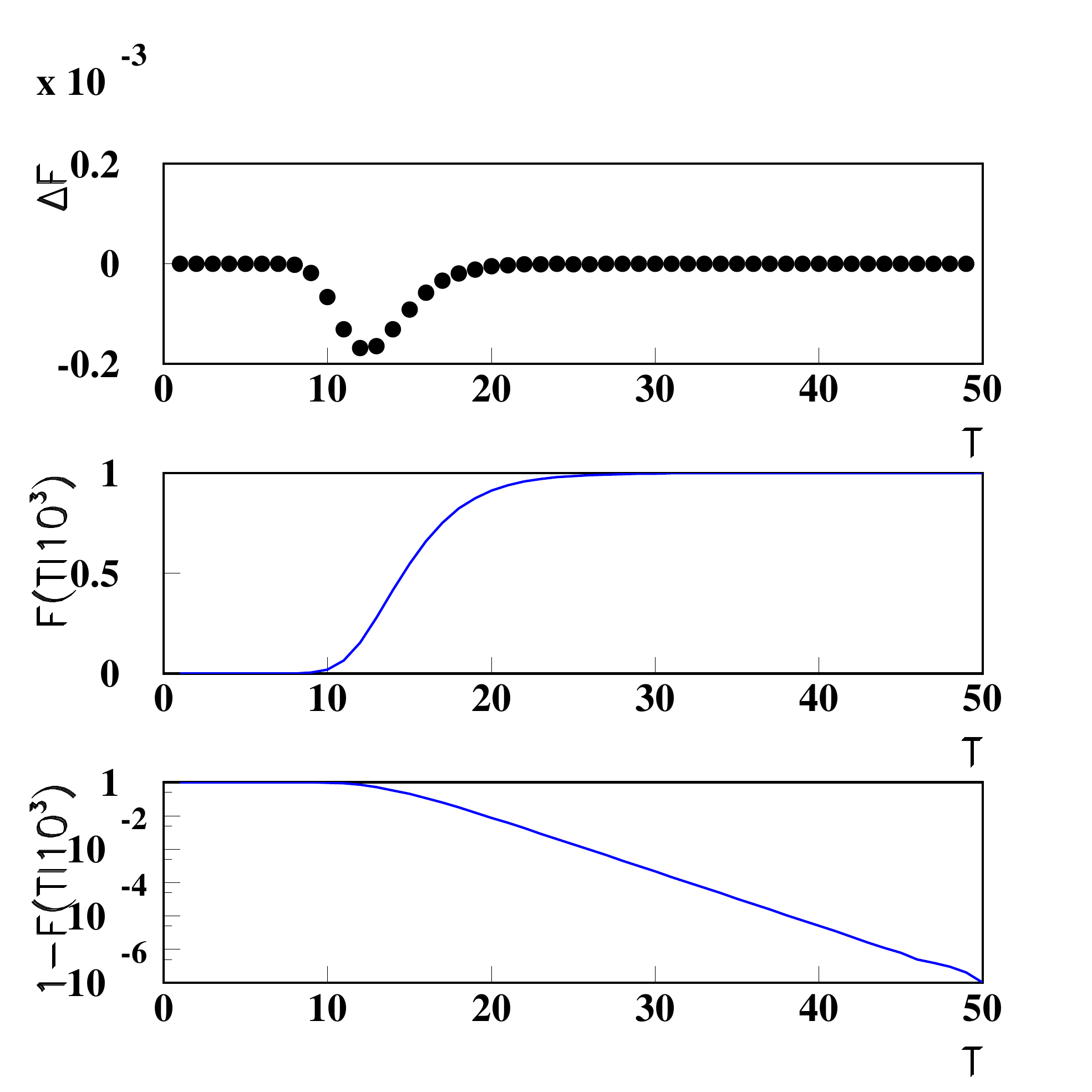}
    \includegraphics[width=2.5in]{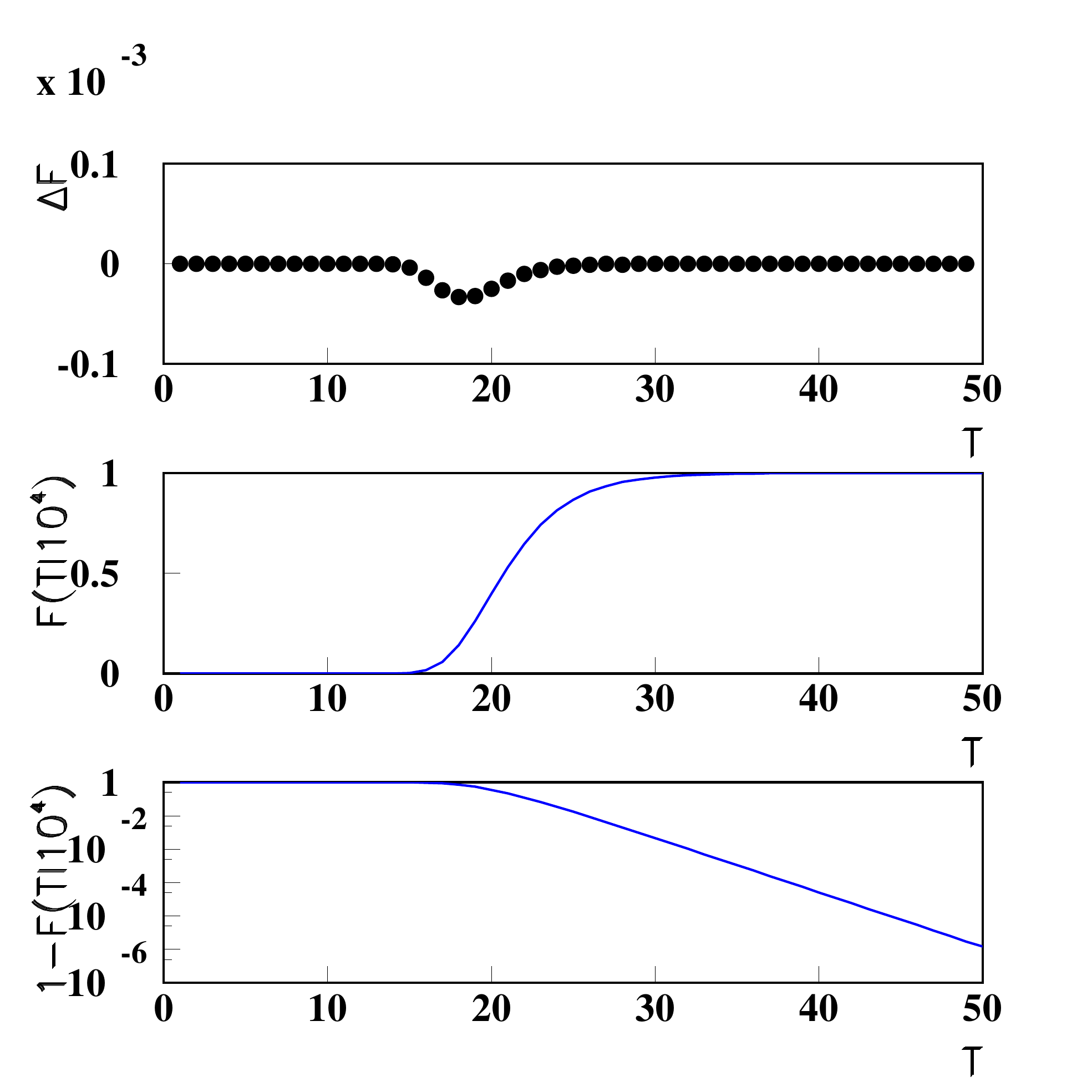}
  \includegraphics[width=2.5in]{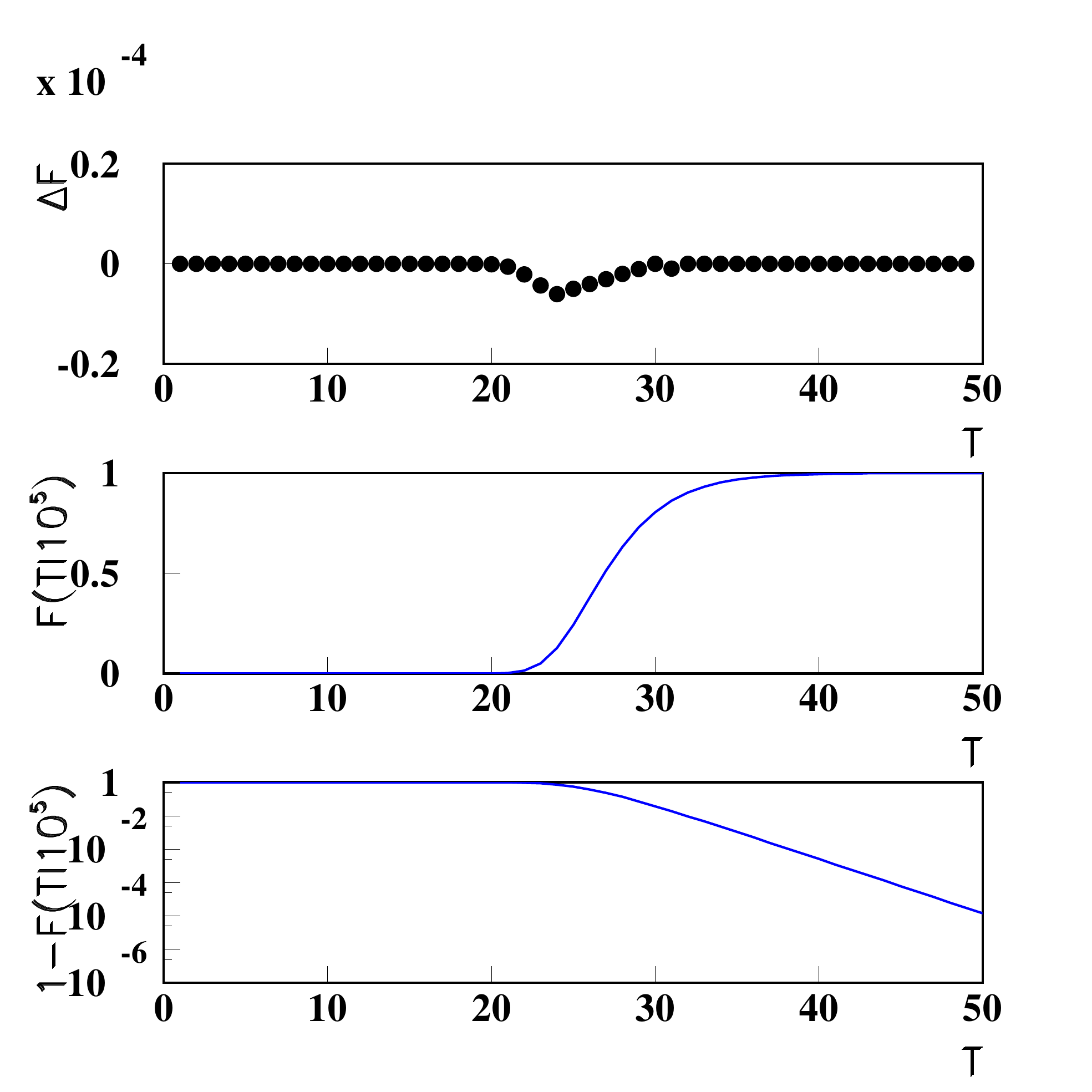}
  \caption{Upper left: the difference between the exact calculation for
    $F(T|100)$ and the scaling equation ~\ref{eq:FTnN} for $F(T|2\cdot 50)$ is
    shown in the upper panel, while the lower panel shows the fractional error
    in the $p$ value as a function of the $p$ value. For the remaining plots, we
    compare $F(T|n\cdot 100)$ with $F(T|2n\cdot 50)$.  Upper right: $n=10$.
    Lower left:$n=100$. Lower right: $n=1000$. The value of $n$ for the $N=50$
    case is always twice that used for $N=100$ so that $L=Nn$ is the same. The
    value plotted in the upper plots is $F(T|n\cdot 100) - F(T|2n\cdot 50)$, the
    second plot shows the cumulative, and the third plot the $p=1-F$ value on
    the log scale.}
   \label{fig:Nncomp}
\end{figure}

We now investigate the size of the uncertainty introduced by our approximations
in deriving our scaling formula. For this purpose, we compare the exact
calculation for $F(T|L=100)$ with the approximation for $N_l=N_r=50$ in
Fig.~\ref{fig:Nncomp} (upper left plots). The top panel shows the difference in
the cumulative probabilities, while the lower panel shows the fractional
difference in the $p$ value as a function of the $p$ value. As is seen in the
figure, our scaling formula gives a very good agreement with the exact
calculation with a maximum difference in the cumulative of about $7\cdot
10^{-4}$.
%\Fred{From the plot, I see $0.07 \cdot 10^{-2} = 7 \cdot 10^{-4}$.  Please check exponent.}
 \noindent It is also found that the difference between the exact
calculation and the approximate calculation decreases as $L$ and $T$ increase.
The fractional error on the $p$ value at small $p$ values is negligible.

To further study the accuracy of expression \ref{eq:FTnN}, we take the
difference of the cumulative distributions, $F(T|n\cdot100) - F(T|2n\cdot50)$
for $n=10, 100, 1000$ and plot these versus the value of the test statistic in
Fig.~\ref{fig:Nncomp}.  We see here that the calculations agree to better than
the per mil level, with the largest differences visible for moderate values of
the cumulative. This is exactly the region where the approximation used in
\refeq{approx} was expected to show small deviations.  The correction term as
evaluated is too small, so that $F(T|2n \cdot 50)>F(T|n \cdot 100)$.  The effect decreases at larger $T$. In the most interesting case, where the $p$ value is small ($F(T)$ large), the calculations are in excellent agreement with negligible differences.

\begin{figure}[htbp] %  figure placement: here, top, bottom, or page
   \centering
  \includegraphics[width=2.5in]{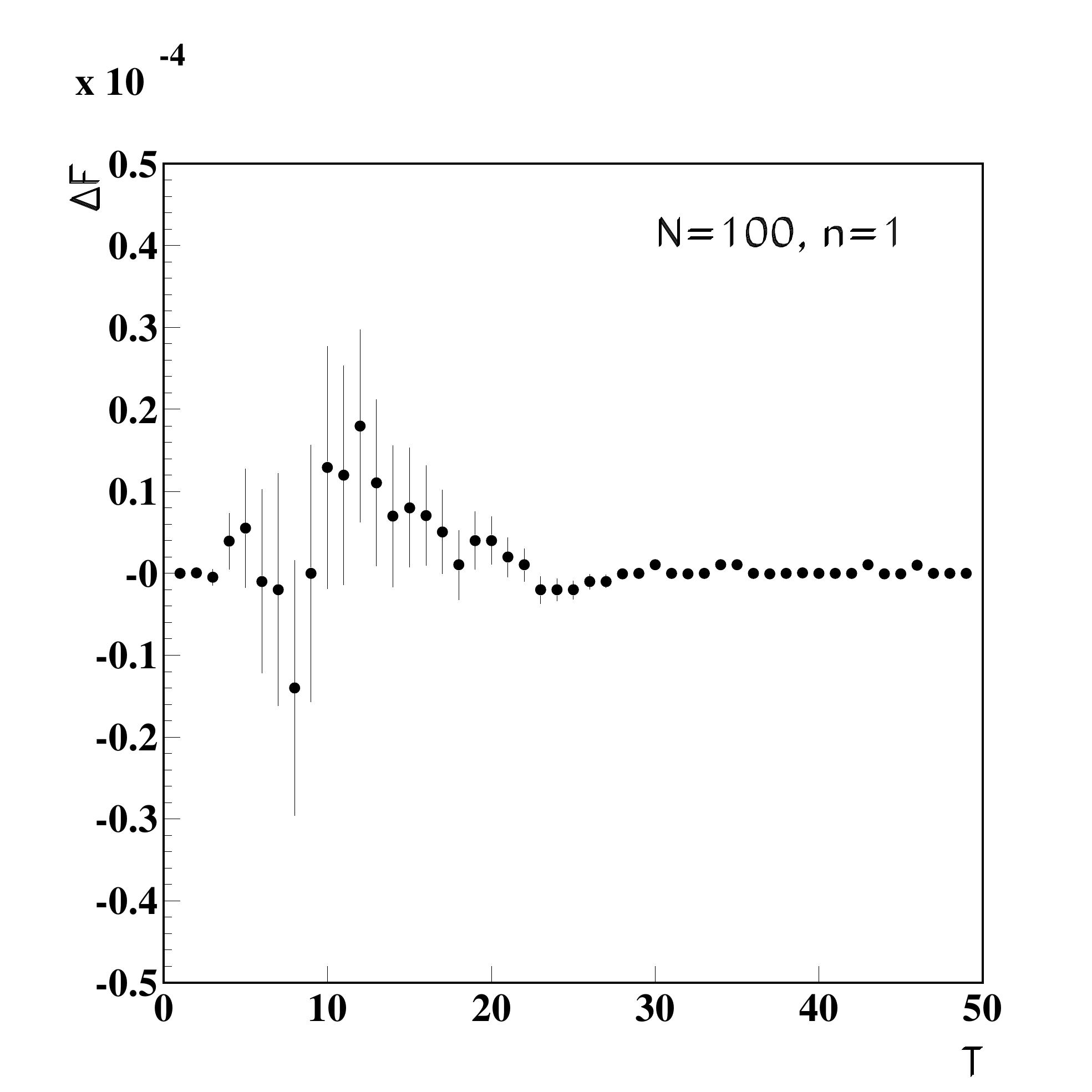}
  \includegraphics[width=2.5in]{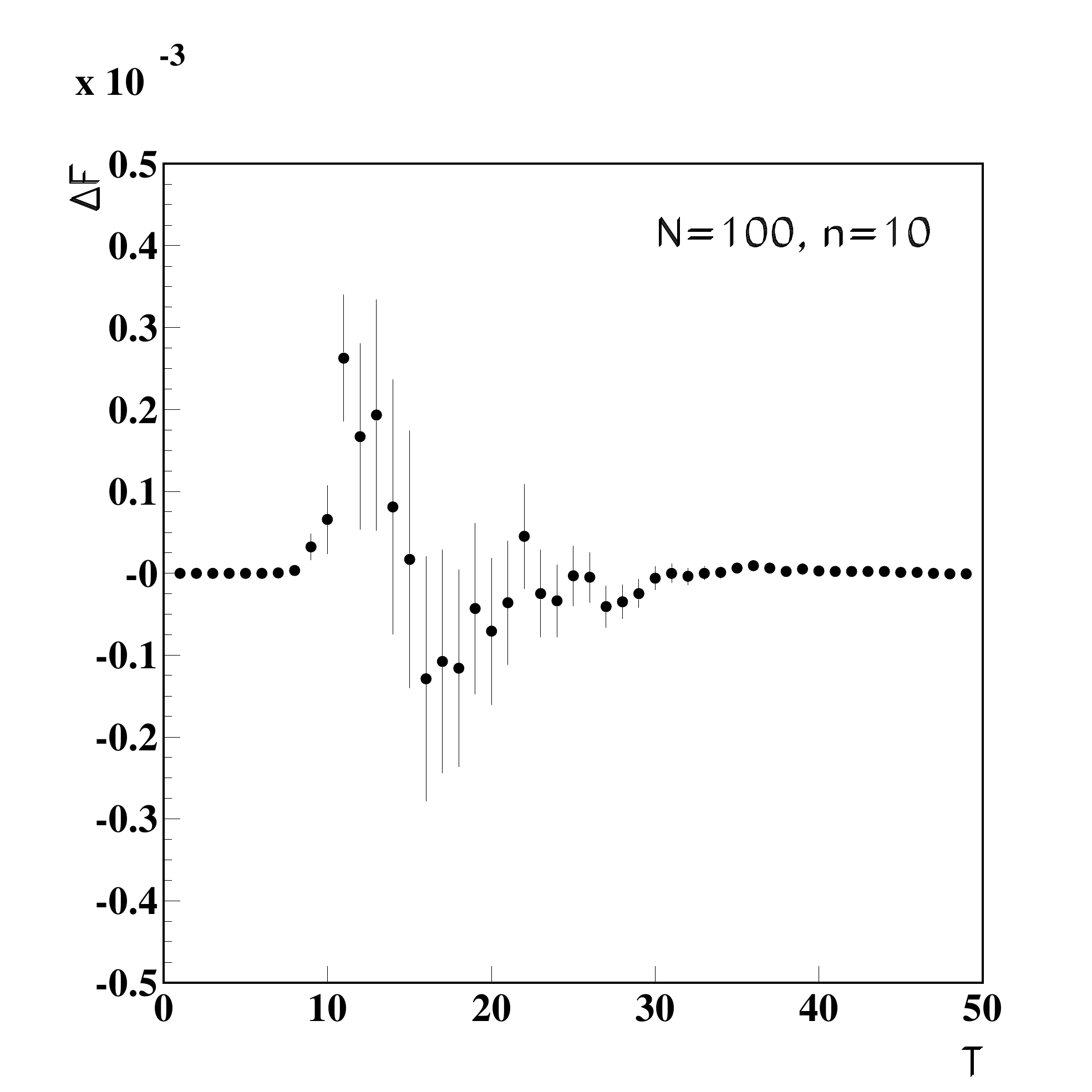}
    \includegraphics[width=2.5in]{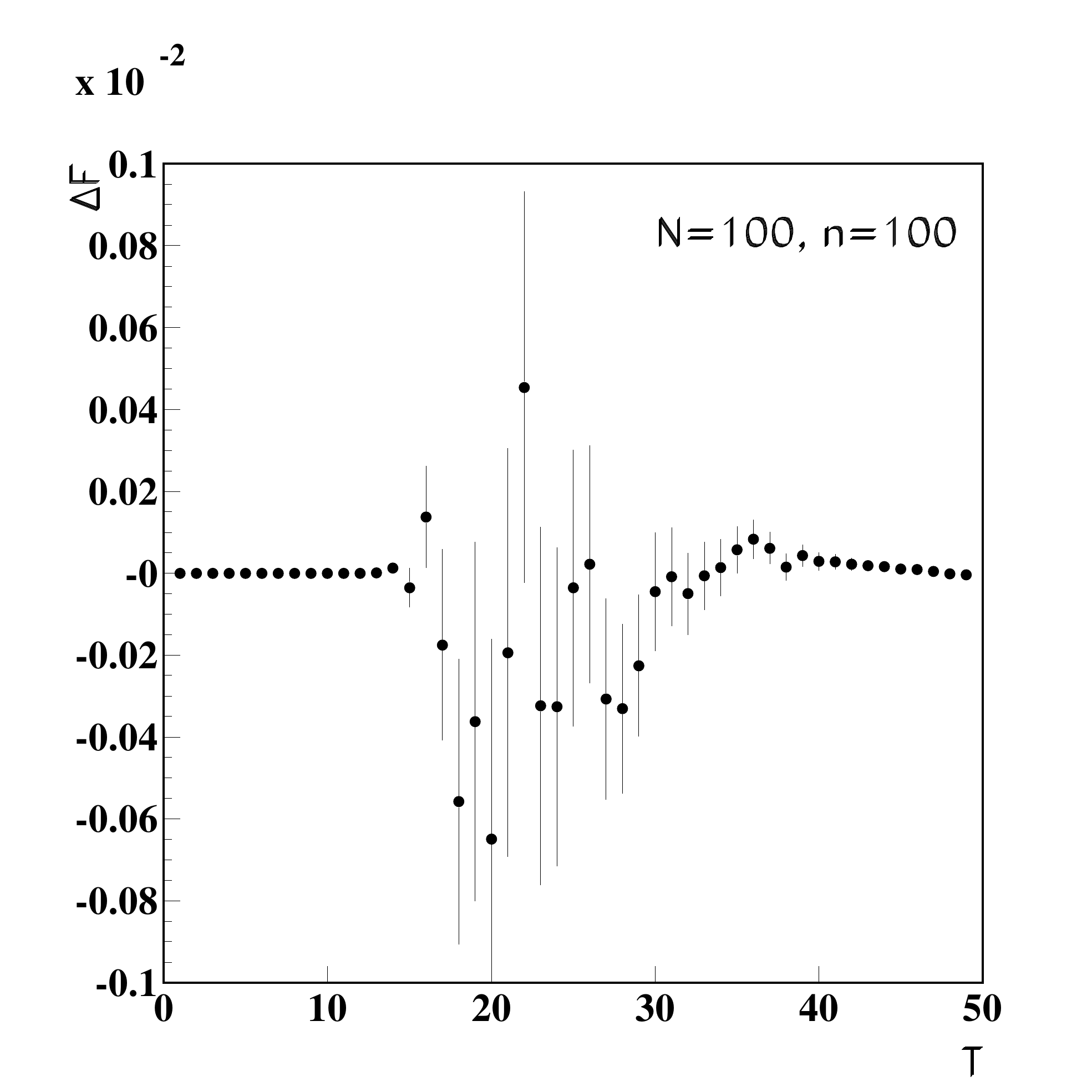}
  \includegraphics[width=2.5in]{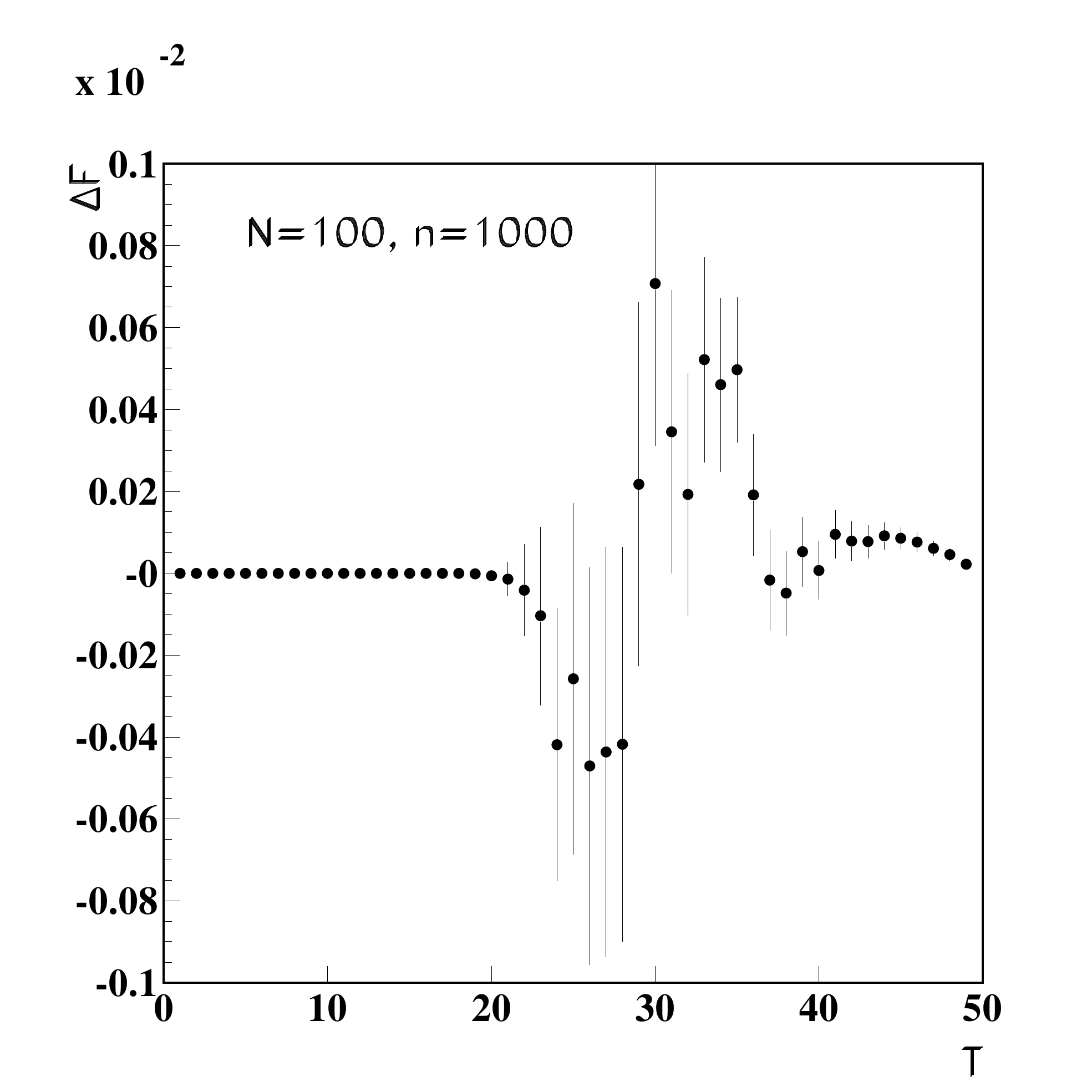}
  \caption{Comparison of Monte Carlo generated cumulative distributions with our
    calculations based on $N=100$.  Upper left: $n=1$. Upper right: $n=10$.
    Lower left: $n=100$. Lower right: $n=1000$. The values plotted are
    $F(T|n\cdot 100) - F_{Monte Carlo}(T|n \cdot 100)$.}
  \label{fig:MCcomp}
\end{figure}

Finally, we compare the calculated cumulative probabilities with results from Monte Carlo simulations using the MT19937 random number generator~\cite{matsumoto1998mersenne} available in the Gnu Scientific Library~\cite{gough2009gnu}.
The Box-Muller algorithm from the Gnu Scientific Library is then used to calculate Gaussian random numbers~\cite{box1958note}. For the Monte Carlo simulations, we generate a large number of experiments with different-length runs, as specified in the table, and keep track of the value of the test statistic in each experiment.  We then form the Monte Carlo cumulative probability and compare to that calculated.  We estimate the statistical uncertainty on the Monte Carlo result using the binomial probability standard deviation~\footnote{Note that the resulting uncertainties are correlated since we are using the observed cumulative for one set of experiments.}. The results are shown in Fig.~\ref{fig:MCcomp}.  For the analytical calculations, we used the $N=100$ set of results.

\begin{table}
\begin{tabular}{ccc}
\hline
$L$ & Experiments & Calculation \\
\hline
$100$ & $10^9$ & Exact analytic calculation \\
$1000$ & $10^7$ & $N=100, n=10$  \\
$10000$ & $10^6$ &  $N=100, n=100$  \\
$100000$ & $10^6$ &  $N=100, n=1000$  \\
\hline
\end{tabular}
\caption{The Monte Carlo data sets generated.  The length of the sequences and the numbers of simulated experiments are given.  The last column gives the parameters used for the calculation.}
\end{table}

For the case $L=100$ we compare the simulation with the random number generator directly with the exact calculation as a test of the quality of the generator.  This test is shown in the upper left panel of Fig.~\ref{fig:MCcomp}.  As can be seen in the figure, the results are within the statistical fluctuations expected from the binomial distribution.  %We note that other random number generators failed this first test.

Given that we have a good random number generator, we can then check the agreement between our calculation given in \refeq{FTnN} for different $n=10,100,1000$.  The tests are shown in the other three panels.  It is seen that in all cases the differences are in agreement with expected statistical fluctuations.

\subsection{Precision of the calculation}
A typical use of our results will be to evaluate the $p$ value for an observed excess, where small $p$ values will generate interest in follow-up analyses.  The $p$ value for the maximum run statistic in a sequence of $L$ measurements is given by
\begin{equation}
p=1-F(T|L) \; .
\end{equation}
Values of $F(T|N)$ for $N\leq 100$ can be calculated exactly to double floating-point precision using the results presented in ~\cite{beaujean2011test, frederik_beaujean_2017_845743}. For $L>N$, we evaluate $F(T|L)$ by finding the
value $n=L/N$ and evaluating $F(T|nN)$ using expression ~\refeq{FTnN} even if $n \notin \mathbb{N}$. For the test case $L=355$, we verified that the three alternatives $(n=3.55, N=100)$, $(n=5, N=71)$, and linear interpolation between $(n=4, N=88)$ and $(n=4, N=89)$ agree to nine significant digits.

The precision can be evaluated as
\begin{eqnarray*}
\dd{F(T|nN)} &=& \frac{nF(T|N)^{n-1} \dd{F(T|N)}}{(1+\Delta(T))^{n-1}}\oplus \frac{(1-n)F(T|N)^n \dd{\Delta(T)}}{(1+\Delta(T))^{n}} \\
&\leq& n \dd{F(T|N)} \oplus (1-n) \dd{\Delta(T)}
\end{eqnarray*}
where the inequality holds since $F(T|N)\leq 1$ and $\Delta(T)\geq 0$, and
$\oplus$ indicates addition in quadrature. To reach a given precision $\epsilon$
on $p$, we require that $F(T|N)$ and $\Delta(T)$ be evaluated to an absolute
precision $\epsilon/n$. As an example, for $L=10^6$, $N=100$, and $\epsilon =
10^{-5}$, we would need an absolute precision on $F(T|N)$ and $\Delta(T)$ at the
$10^{-9}$ level. We have verified that this can be reached in practice:
regarding $F(T|N)$, our two implementations \cite{frederik_beaujean_2017_845743} in \texttt{mathematica} and
\texttt{C++} agree at the $10^{-15}$ level, and with 1D numerical integration $\Delta(T)$
can be computed at the $10^{-10}$ level.

\section{Test case - (Fake) axion search}

\begin{figure}[htbp] %  figure placement: here, top, bottom, or page
   \centering
  \includegraphics[width=6in]{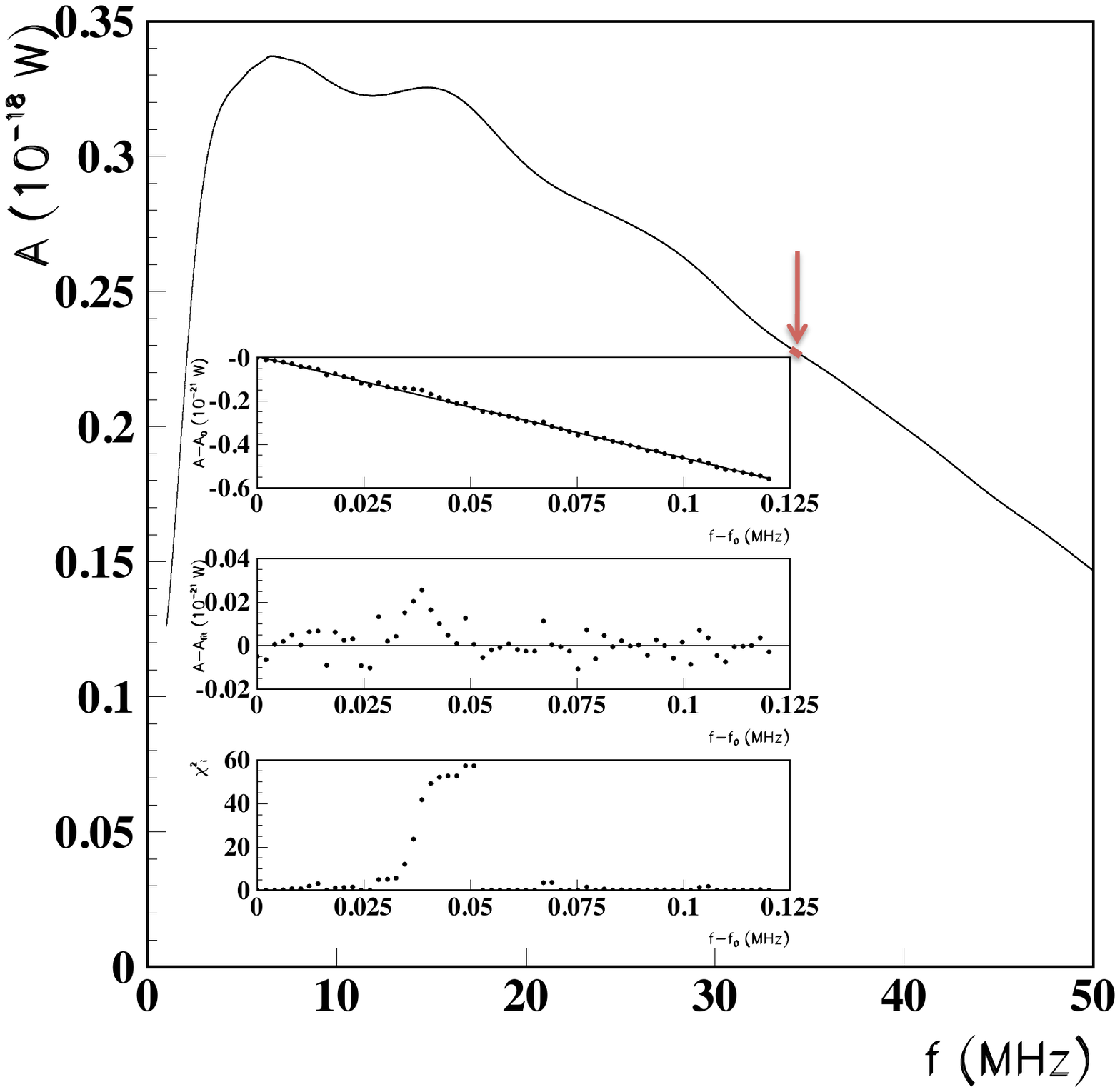}
  \caption{The power spectrum to be analyzed. The data are from a test setup for an axion search experiment developed at the Max Planck Institute for Physics. A signal corresponding to a possible axion signal was introduced in the setup and searched for using the test statistic described in this paper. The region of the signal is shown in the inset together with the background fit (top inset), the residuals from the fit (middle inset) and the run $\chi^2$ (lower inset).}
  \label{fig:Spectrum}
\end{figure}

As an example of the use of our run statistic, we consider an experimental setup
at the Max Planck Institute for Physics designed to search for axions in the
mass range $40-400~\mu$eV~\cite{TheMADMAXWorkingGroup:2016hpc}. The measurement
is effectively a power spectrum as a function of the frequency of emitted
microwave radiation built from many independent measurements. The baseline
signal is dominated by the thermal background ($10^{-19}$~W). A weak fake axion
signal was injected; the location and width of the signal were unknown when the
analysis was carried out. Although the shape (Gaussian) was known, this
information was not used. The spectrum to be analyzed is shown in
Fig.~\ref{fig:Spectrum}. The shape of the background spectrum and level of
fluctuation was unknown, and had to be determined from the data by assuming that
any possible signal would provide a much sharper feature than any change in
background. In the example considered here, the spectrum consists of 24576 data
points giving the integrated power in $\approx 2$~kHz intervals. A signal is
expected to be one or a few bins wide. To minimize the number of assumptions
made about the signal shape, we scanned the spectrum with our run statistic to
determine if there was a significant deviation from background expectations.

To determine the background shape and fluctuations, the full spectrum was
partitioned into contiguous subsets and a second order polynomial fit was used
to find the background shape. The residuals were used to extract the standard
deviation of the fluctuations. Different lengths of the subsets were considered
from a minimum of 96 measurements to 256 measurements. It was verified that the
fluctuations relative to the fit function followed the expected Gaussian
distribution.

\begin{figure}[ptb] %  figure placement: here, top, bottom, or page
   \centering
  \includegraphics[width=3in]{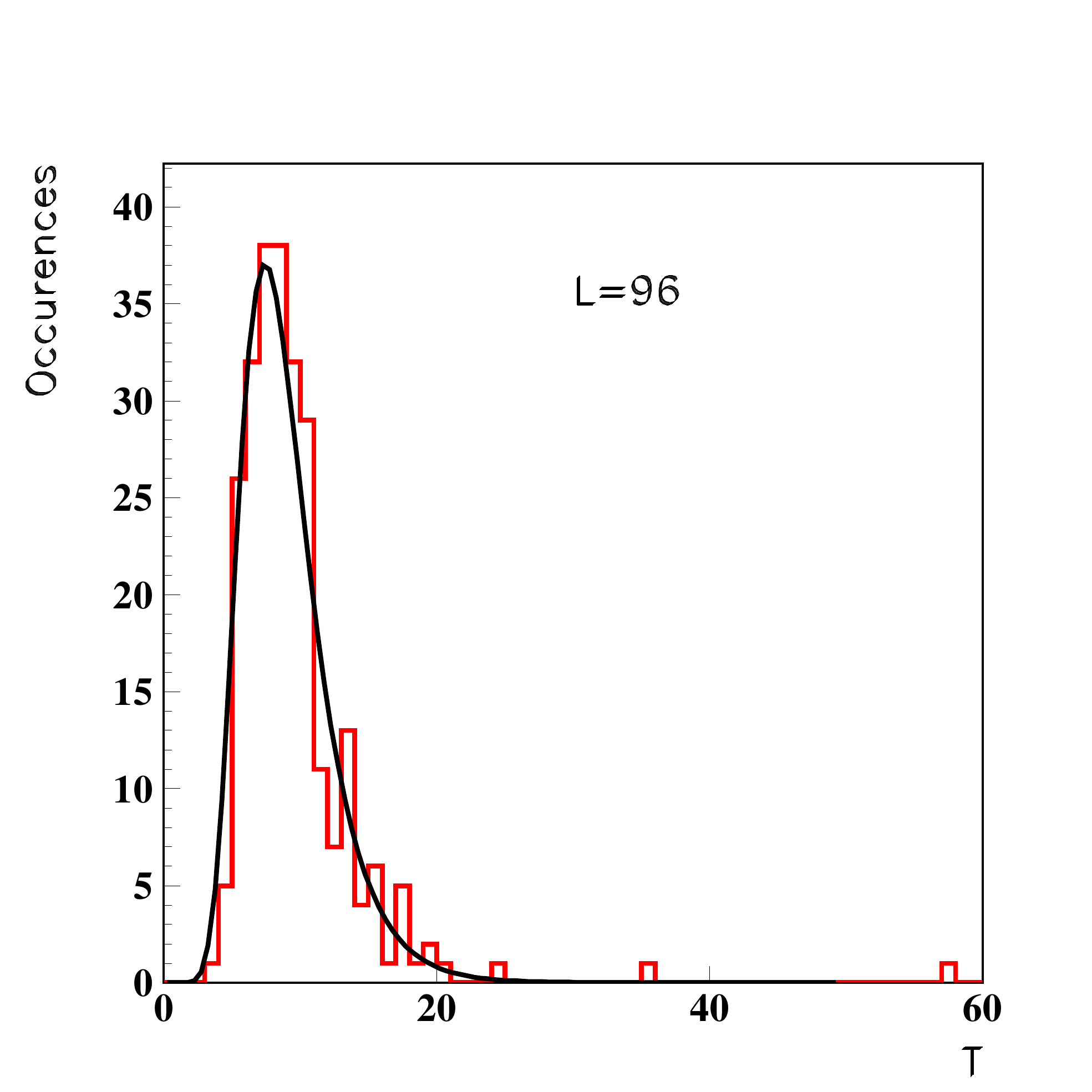}
  \caption{The distribution of the test statistic calculated in the 256 sets of
    96 samples, compared to the expected distribution for $L=96$.}
  \label{fig:TS}
\end{figure}

The distribution of $T$ for the 256 sets of 96 data points in shown in
Fig.~\ref{fig:TS}, and compared to the expectation from the exact calculation
of~\cite{beaujean2011test} for $N=96$. The largest value found was $\Tobs =
57.3$. For $N=96$, that is if the total number of observations had been only
$96$, this has a $p$ value of $p=6.4\cdot 10^{-9}$. To get the $p$ value for all
observations, we use the expression \refeq{FTnN} and find a $p$ value of $p=1.9
\cdot 10^{-6}$, which is very small. The frequency at which this signal was
found was indeed the frequency of the injected 'fake axion'. For comparison, the
second most significant $p$ value of a test statistic found in one of our runs
was $2 \cdot 10^{-5}$ taking $L=96$, which becomes $p\approx 6 \cdot 10^{-3}$
when taking the full $24576$ samples into account. These changes of the $p$
value illustrate the importance of the look-elsewhere effect. To infer the $p$
value from $N$ for $nN$ observations, the obvious guess for the trials factor in
\refeq{trials-overall} is 256, which is too small by 20 \% for the above
numbers. Our more accurate result based on \refeq{FTnN} is achieved for
essentially the same computational effort.

Figure~\ref{fig:Spectrum} shows in the inset the data in the range of the
candidate signal together with the polynomial background fit, the residuals, and
the running $\chi^2$ value as a function of the frequency.  The test was repeated several times with different fake signals and these were found in every case.

To appreciate the usefulness of \refeq{FTnN}, we consider the
computational cost. Proposition 1 from \cite{beaujean2011test} states
that computing the exact expression for the distribution of $T$ for
$L=24576$ requires a sum over the enormous number of
$2.6 \cdot 10^{169}$ integer partitions, something that is completely
unfeasible with the best supercomputers today. But with our
approximate result, we only need to compute the exact expression for
$N=96$ which requires $1.4 \cdot 10^8$ partitions, or a reduction of
work by 161 orders of magnitude. The algorithm is of the streaming
type and the partitions need not be stored but can be independently
processed, so it makes ideal use of modern multi-core computers. On a
desktop computer with an Intel i7-4700 CPU with four cores and eight
\texttt{openMP} threads, the \texttt{C++}
implementation~\cite{frederik_beaujean_2017_845743} of $F(T|N=100)$ is
computed in $1.8$ s. With some reduction in precision and range of
validity, computing for $N=50$ as the baseline requires less than 10 ms.

\section{Conclusion} \label{sec:conclusion}

We presented an extension of our previous work~\cite{beaujean2011test}
where we introduced the runs statistic $T$ to detect bumps that are
incompatible with a background model for ordered 1D data sets such as
spectra. We derived the exact distribution of the statistic needed to
compute a $p$ value before but the expression could not be evaluated
beyond about 100 data points within reasonable time. In this work, we
describe an approximation \refeq{FTnN} that takes the results from few
data points and extrapolates to millions of points in a principled
manner by dividing the data into chunks. In the region of interest
where the observed value $\Tobs$ is large and the $p$ value is small,
the approximation has both high accuracy and precision. We recommend
to use the exact expression for $100$ data points as the basis for
extrapolation but in most applications one may even start lower
without loss of precision. The largest discrepancies between exact and
approximate distribution appear for intermediate values of $T$ but
even there they are too small to change the judgement of the quality
of the model. The code implementing all expressions in this paper is
avalailable online at \url{https://github.com/fredRos/runs}.

Through Monte Carlo experiments we validated our results up $10^5$
data points. In addition, the test statistic is computed for a
real-life physics problem of detecting a fake axion signal in a power
spectrum of 24576 data points and shows a very significant excess from
the background model at the expected location. All other properties
studied in this example are in full agreement with our
derivations. This provides the basis to apply this method when the
axion experiment has grown to full scale with milllions of data
points.

% \begin{acknowledgments}
% Thanks to \dots
% \end{acknowledgments}

\bibliography{references}

\end{document}